\documentclass[aps,amsmath,amssymb,preprint,showkeys,superscriptaddress,showpacs,floatfix,longbibliography]{revtex4-1}
\usepackage{graphicx}
\usepackage{dcolumn}
\usepackage[colorlinks=true,linkcolor=blue,citecolor=blue,urlcolor=blue]{hyperref}
\usepackage{multirow}
\usepackage[usenames,dvipsnames]{xcolor}
\usepackage{soul}
\usepackage{braket}
\usepackage{bm}
\usepackage{nicefrac}
\usepackage{siunitx}
\DeclareSIUnit\rydberg{Ry}
\usepackage{xcolor}
\usepackage[utf8]{inputenc}
\usepackage{upgreek}
\usepackage{tikz}

\begin{document}

\title{Trends in the hyperfine interactions of magnetic adatoms on thin insulating layers}

\author{Sufyan Shehada}\email{s.shehada@fz-juelich.de}
\affiliation{Peter Gr\"{u}nberg Institut and Institute for Advanced Simulation, Forschungszentrum J\"{u}lich and JARA, 52425 J\"{u}lich, Germany}
\affiliation{Department of Physics, RWTH Aachen University, 52056 Aachen, Germany}
\author{Manuel dos Santos Dias}
\affiliation{Peter Gr\"{u}nberg Institut and Institute for Advanced Simulation, Forschungszentrum J\"{u}lich and JARA, 52425 J\"{u}lich, Germany}
\author{Filipe Souza Mendes Guimar\~{a}es}
\affiliation{J\"ulich Supercomputing Centre, Forschungszentrum J\"{u}lich and JARA, 52425 J\"{u}lich, Germany}
\author{Muayad Abusaa}
\affiliation{Department of Physics, Arab American University, Jenin, Palestine}
\author{Samir Lounis}
\affiliation{Peter Gr\"{u}nberg Institut and Institute for Advanced Simulation, Forschungszentrum J\"{u}lich and JARA, 52425 J\"{u}lich, Germany}
\affiliation{Faculty of Physics, University of Duisburg-Essen, 47053 Duisburg, Germany}

\date{\today}

\begin{abstract}
 Nuclear spins are among the potential candidates prospected for quantum information technology. A recent breakthrough enabled to atomically resolve their interaction with the electron spin, the so-called hyperfine interaction, within individual atoms utilizing scanning tunneling microscopy (STM). Intriguingly, this was only  realized for a few species put on a two-layers thick MgO. Here, we systematically quantify from first-principles the hyperfine interactions of the whole series of 3d transition adatoms deposited on various thicknesses of MgO, NaF, NaCl, h--BN and Cu$_2$N films. We identify the adatom-substrate complexes with the largest hyperfine interactions and unveil the main trends and exceptions. We reveal the core mechanisms at play, such as the interplay of the local bonding geometry and the chemical nature of the thin films, which trigger transitions between high- and low-spin states accompanied with subtle internal rearrangements of the magnetic electrons. By providing a general map of hyperfine interactions, our work has immediate implications in future STM investigations aiming at detecting and realizing quantum concepts hinging on nuclear spins.
\end{abstract}

\maketitle

\begin{figure}[tb]
    \centering
    \includegraphics[width=\textwidth]{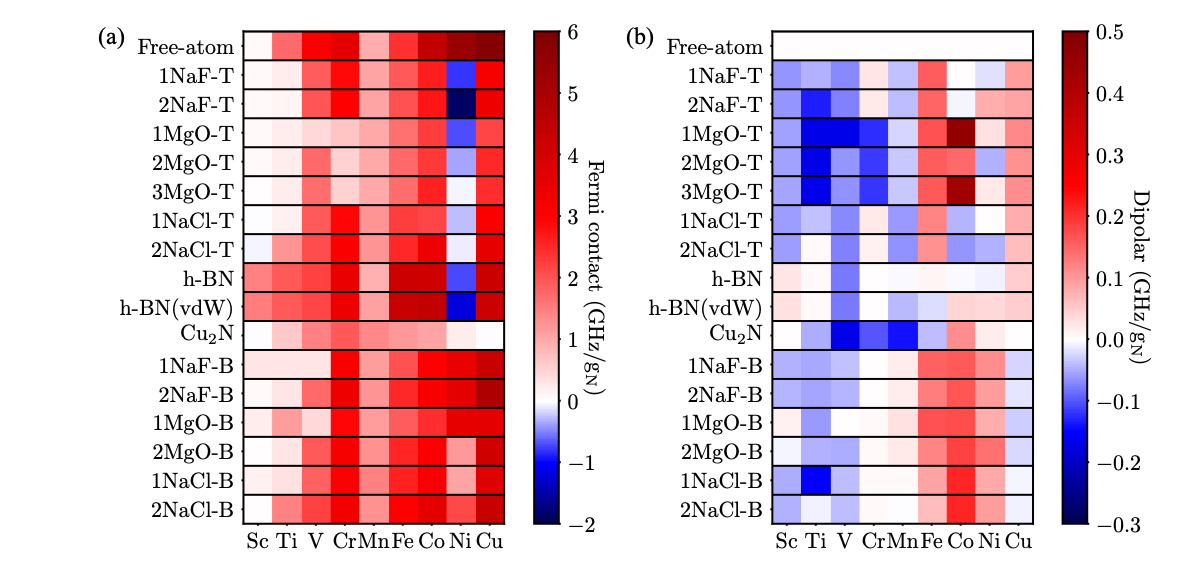}
    \caption{Hyperfine interactions of magnetic adatoms on ultrathin films:
    (a) Fermi contact and (b) dipolar contributions.
    The free-atom values are given for comparison.
    The number to the left in the legend indicates the number of layers in the film.
    The adsorption position is denoted with T for atop the anion and with B for the bridge position (for h--BN and Cu$_2$N only T was considered).
    The possible impact of corrections from van der Waals interactions was explored for the h--BN case.}
    \label{fig:all_mag_results}
\end{figure}

\section{Introduction}

Magnetic atoms on surfaces have attracted a lot of attention due to potential applications in magnetic storage and quantum computation~\cite{Choi2019a}.
The coupling between the magnetic atoms and the surface leads to challenges and opportunities.
This coupling is a necessary ingredient in order to encode information in the atomic magnetic moment, but at the same time leads to its destabilization, so information may be lost~\cite{Delgado2017a}.
Two main approaches have so far been pursued to overcome these difficulties: to employ thin insulating layers between the adatoms and substrate to suppress the interactions with the substrate's conduction electrons~\cite{Hirjibehedin2006,Hirjibehedin2007,otte2008role,Loth2010,Loth2010a,Loth2012,PhysRevLett.109.066101,PhysRevLett.112.026102,Rau2014,Baumann2015,jacobson2015quantum,Donati2016,natterer2017reading,Paul2017,Yang2017,PhysRevB.100.180405,gallardo2019large,Rejali2020}, or to bring several magnetic atoms together to reduce their quantum mechanical fluctuations~\cite{Khajetoorians2011a,Khajetoorians2011,Khajetoorians2012,Khajetoorians2013,Hermenau2017}.
Experimental investigations of magnetic adatoms on thin insulating layers have mostly centered on Cu$_2$N ~\cite{Hirjibehedin2007,Hirjibehedin2006,otte2008role,Rejali2020,Loth2010,Loth2010a,Loth2012} and MgO~\cite{Rau2014,Baumann2015,Paul2017,Yang2017,Donati2016,natterer2017reading,PhysRevB.100.180405}, and to a lesser extent on h--BN~\cite{PhysRevLett.109.066101,jacobson2015quantum,gallardo2019large}, usually in connection with scanning tunnelling microscopy (STM)~\cite{Wiesendanger2009}.

It has also been suggested that the nuclear magnetic moment instead of the electronic one could be used for application purposes.
The two are coupled via the hyperfine interaction~\cite{Slichter:801180}, which provides insight into the electronic structure and chemical bonding of atoms, molecules and solids, as explored with nuclear magnetic resonance techniques. 
Individual nuclear spin states tend to have a very long lifetime and hold great promise as building blocks for quantum computers (qubits)~\cite{blinov2004quantum,Thiele1135,vincent2012electronic}. 
Two promising steps in this direction were recently achieved experimentally through the development of single-atom electron paramagnetic/spin resonance (EPR/ESR)~\cite{Balatsky2012,Baumann2015a,Choi2017a,Yang2017,Willke2018,Willke2018a,Bae2018,yang2018electrically,Yang2019a,Willke2019a,Seifert2020}:
The first one is the recent detection of the hyperfine interaction between the atomic nucleus and surrounding electrons for single Fe and Ti adatoms on MgO/Ag(001)~\cite{Willke2018a};
The second is the control of the nuclear polarization of individual Cu adatoms on the same surface~\cite{yang2018electrically}.
The actual mechanism underpinning the EPR/ESR experiments is still under investigation~\cite{Caso2014,Berggren2016,Lado2017,Reina-Galvez2019,Seifert2020,doi:10.1021/acs.jpca.9b10749}, but it is clear that a deep understanding of the hyperfine interactions in these systems is an essential piece of the puzzle. Moreover, the interactions were detected intriguingly only for a few set of atoms on a two-layers thick MgO film. 

In this work, we report on vast systematic first-principles calculations of the hyperfine interactions for magnetic transition metal adatoms (from Sc to Cu) placed on different ultrathin insulators with different thickness and bonding site, namely MgO, NaF, NaCl, h--BN and Cu$_2$N. 
A summary of our findings is given in Fig.~\ref{fig:all_mag_results} for both contributions to the hyperfine interactions, which are measurable experimentally~\cite{Willke2018a}: the Fermi contact and dipolar terms. 
This enables the identification of the adatom-substrate complexes with the largest interactions and those that can even experience a sign change of the interaction, translating to an antiferromagnetic nuclear-electron spin coupling.
The obtained map of interactions can serve as a guide for future experimental explorations of nuclear-electron spin physics of adatoms and complex nanostructures on surfaces.

We analyze the trends and the dependence of the computed hyperfine parameters on the filling of the magnetic $d$-orbitals of the adatom and on the type and strength of the bonding with the substrate, on the thickness of the ultrathin film (for MgO, NaF and NaCl), and on two choices of binding sites. 
We show how the hyperfine interactions give access to information about the local electronic structure and what are the main quantities that determine their properties.

\section{Results}

\subsection{The hyperfine interaction}
The hyperfine interaction is the interaction between the electron spin  $\mathbf{S}$ and the nuclear spin $\mathbf{I}$, and can be described by using the following hyperfine Hamiltonian~\cite{PhysRevB.85.115430,PhysRevB.47.4244}:
\begin{equation}
    \hat{H}= \mathbf{S}\cdot\underline{\mathrm{A}}(\mathbf{R})\cdot\mathbf{I} \;,
\end{equation}
where $\underline{\mathrm{A}}(\mathbf{R})$  is the hyperfine coupling tensor associated with the nucleus located at position $\mathbf{R}$, and the angular momenta are measured in units of $\hbar$.
It consists of two contributions~\cite{PhysRevB.47.4244}, $\mathrm{A}_{ij}= \mathrm{a}\,\delta_{ij}+\mathrm{b}_{ij}$ ($i,j=x,y,z$).
Both quantities ($\mathrm{a}$ and $\mathrm{b}$) depend on the prefactor $P = \mu_0 g_\mathrm{e} \mu_\mathrm{B} g_\mathrm{N} \mu_\mathrm{N}$, with $\mu_0$ the vacuum permeability, the electron and nuclear g-factors, $g_\mathrm{e}$ and $g_\mathrm{N}$, and the Bohr and nuclear magnetons, $\mu_\mathrm{B}$ and $\mu_\mathrm{N}$.
We discuss these two contributions without accounting for the effects of the spin-orbit interaction, which is consistent with the scalar-relativistic approximation adopted in our calculations.

The first contribution is called the Fermi contact term,
\begin{equation}\label{Fermi}
    \mathrm{a}= \frac{2P}{3}\,\rho_\mathrm{s}(\mathbf{R}) \;.
\end{equation}
It provides the isotropic part of the hyperfine interaction and originates from a finite electron spin density   $\rho_\mathrm{s}(\mathbf{r})=\rho_\uparrow(\mathbf{r})-\rho_\downarrow(\mathbf{r})$ at the position of the nucleus.
This is usually the dominant contribution.

The second contribution is from the dipolar interaction between the electron and nuclear spin,
\begin{equation}\label{Dipolar}
    \mathrm{b}_{ij}= \frac{P}{4\pi} \int\mathrm{d}\mathbf{r}\;\frac{3r_{i}r_j-r^2\delta_{ij}}{r^5}\,\rho_\mathrm{s}(\mathbf{r}) \;.
\end{equation}
Here $\mathbf{r}$ is relative to the nuclear position $\mathbf{R}$ and $r=|\mathbf{r}|$.
The tensor $\mathrm{b}_{ij}$ is traceless and has the angular symmetry of a quadrupole-like moment of the spin density, although heavily weighted towards the nucleus.
It thus depends on the shape of the spin density and reflects the symmetry of the local atomic environment.
Choosing the local cartesian axes to align with the principal axes of this tensor, we can discuss it in terms of its eigenvalues $\mathrm{b}_{xx}$, $\mathrm{b}_{yy}$ and $\mathrm{b}_{zz} = -(\mathrm{b}_{xx}+\mathrm{b}_{yy})$.

In this work, we report the calculated Fermi contact and dipolar hyperfine interaction per nuclear g-factor, making them applicable for all isotopes of the 3$d$ elements.
For easier comparison with experimental measurements of these quantities, we convert to frequency units (e.g.\ $\mathrm{GHz}$) using Planck's constant.

\subsection{Reference atomic calculations}
We first present the electronic and magnetic properties and the hyperfine parameters for isolated 3$d$ atoms (Sc--Cu), i.e., without any substrate.
These results provide a clear picture of the origin of the hyperfine interaction within our computational approach, and serve as benchmarks to understand the behavior of the same atoms when placed on different ultra-thin insulating layers.

\begin{figure}[tb]
    \centering
    \includegraphics[width=\textwidth]{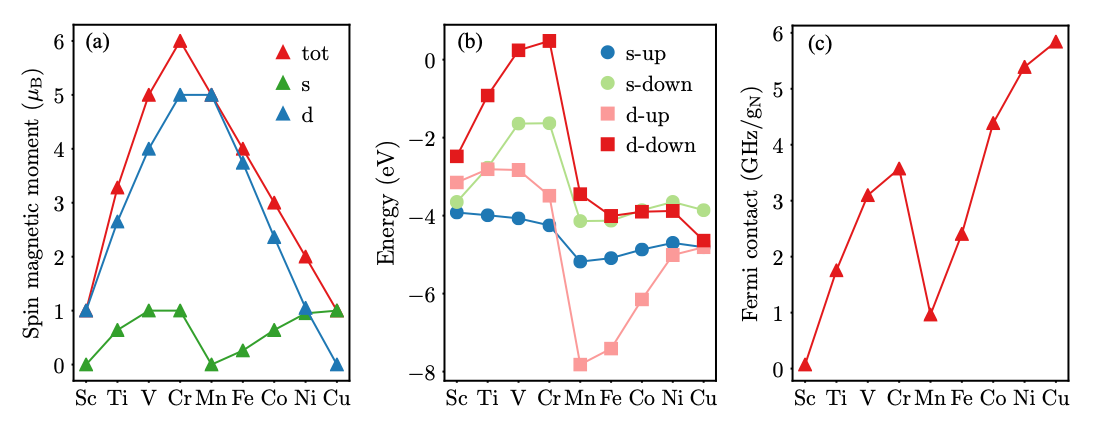}
    \caption{Basic properties of isolated $3d$ transition metal atoms.
    (a) Spin magnetic moment, (b) spin-polarized energy levels and (c) Fermi contact contribution to the hyperfine interaction of isolated $3d$ transition metal atoms.}
    \label{fig:atoms_mag}
\end{figure}

The most fundamental magnetic property of an isolated atom is its electron spin magnetic moment ($\mu_\mathrm{s} = 2\mu_\mathrm{B}\mathbf{S}/\hbar$), shown in  Fig.~\ref{fig:atoms_mag}a.
As expected, we find that the value of the spin moment follows mostly the behaviour from Hund's first rule, with the well-known exceptions of Cr and Cu, as well as two additional cases of Ti and V. 
These exceptions are explained by the spin-polarized atomic energy levels in Fig.~\ref{fig:atoms_mag}b.
Cr and Cu are the famous exceptions to Hund's first rule known from atomic physics.
In the case of Cr, we find that the $s$-up and $d$-up energy levels are lower than the $s$-down and $d$-down, which stabilizes the electronic configuration [Ar]$4s^13d^5$, corresponding to the increase in stability coming from the half-filled $d$-shell.
For Cu we see that the $s$-down level is higher in energy than all the other ones and leads to the electronic configuration [Ar]$4s^13d^{10}$, where the full $d$-shell is known to be more stable than a full $s$-shell.
Ti and V are computational exceptions which have no parallel in atomic physics, and illustrate potential difficulties with the computational approach for the isolated cases.
The ordering of the computed atomic energy levels for V is the same as for Cr, favoring the [Ar]$4s^13d^4$ electronic configuration (see Fig.~\ref{fig:atoms_mag}b).
The properties of V impurities have already been shown to be sensitive to the exchange-correlation functional~\cite{Weber_1997}, with a spin magnetic moment of $5\,\mu_\mathrm{B}$ within the local spin density approximation (LSDA) and $3\,\mu_\mathrm{B}$ for generalized gradient approximation (GGA) calculation. In our case, this sensitivity can also depend on the chosen pseudopotential.
Ti illustrates what happens if two energy levels become nearly degenerate in energy (closer than the employed smearing for the occupations).
Its computed spin moment of $\mu_\mathrm{s} = 3.28\,\mu_\mathrm{B}$ is the consequence of distributing three electrons among the $d$-up and the $s$-down levels due to this near-degeneracy. Despite the issues encountered for two isolated atoms, it is still instructive to analyse the overall trend of the respective hyperfine interaction as function of the atomic magnetic moments.

Due to the spherical symmetry of the free atoms, only the Fermi contact contribution of the hyperfine interaction is finite, and their values are shown in Fig.~\ref{fig:atoms_mag}c.
Its magnitude increases steadily as we go from Sc to Cr, then drops suddenly for Mn and returns to steadily increasing until Cu.
To rationalize this behavior we have to decompose the total spin moment of the free atom into contributions coming from its $s$ and its $d$ electrons, as is also shown in Fig.~\ref{fig:atoms_mag}a.
We see that the Fermi contact contribution follows the behavior of the $s$ contribution to the spin moment~\cite{Akai1990}, and so is a useful experimental probe of this otherwise hard to measure quantity.
Eq.~\eqref{Fermi} shows that the relevant quantity is the spin density at the nuclear position, $\rho_\mathrm{s}(\mathbf{R})$, which arises solely from the $s$ electrons.
This immediately explains the drop in the Fermi contact contribution for Mn, as the $s$-shell becomes full and so $\rho_\mathrm{s}(\mathbf{R})$ becomes strongly reduced in comparison to Cr.
However, a full $s$-shell does not mean a vanishing Fermi contact contribution, as the $s$-up and $s$-down wave functions have a different spatial dependence due to the exchange splitting of the atom, and so will not lead to a full compensation in $\rho_\mathrm{s}(\mathbf{R})$.
This mechanism also applies to the spin-polarized $s$ core electrons.
Furthermore, the $s$ wave functions contract towards the nucleus as the atomic number increases, which explains the overall trend for the Fermi contact contribution to increase when going from Sc to Cu.

\subsection{Adatoms on 1--3 layers of MgO}
We now turn our investigation to magnetic elements deposited on a few layers of MgO.
We report on adatoms that have been experimentally investigated with STM~\cite{Rau2014,Baumann2015,Paul2017,Yang2017,Baumann2015a,Choi2017a,Yang2017,Willke2018,Bae2018,Yang2019a,Willke2018a,Willke2019a,Seifert2020,yang2018electrically}, as well as others for which no measurements have been reported so far.
We point out that the hyperfine interaction was only measured on Ref.~\citenum{yang2018electrically} for Cu (\SI{1.95}{\giga\hertz}/$\mathrm{g_N}$), and on Ref.~\citenum{Willke2018a} for Fe (\SI{1.304}{\giga\hertz}/$\mathrm{g_N}$) and hydrogenated Ti (with two reported values: \SI{0.061}{\giga\hertz}/$\mathrm{g_N}$ on the top of oxygen position, and \SI{0.161}{\giga\hertz}/$\mathrm{g_N}$ on the bridge position). These measurements of the hyperfine interaction were obtained on a substrate composed of two layers of MgO/Ag(001) surface.

\begin{figure}[tb]
    \centering
    \includegraphics[width=\textwidth]{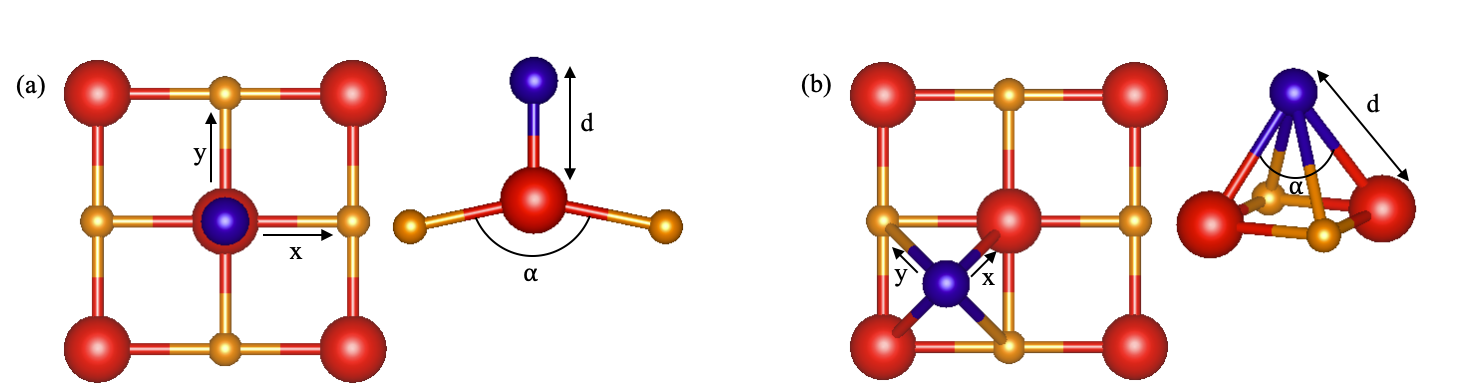}
    \caption{Geometry of adatoms on MgO, NaF and NaCl.
    Top and side views of (a) adatom stacked on top of anion and (b) adatom in the bridge position.
    The adatom is represented by a blue sphere, the anion by a red sphere and the cation by an orange sphere.
    The definition of the distance $d$ between the adatom and anion and the cation-anion-cation bond angle $\alpha$ is also illustrated.}
    \label{fig:mgo_geom_sketch}
\end{figure}

We start by discussing the considered structures and relaxed geometries: the adatoms sit either on top of an oxygen with up to three layers of MgO (Fig.~\ref{fig:mgo_geom_sketch}a), or in a bridge position with up to two MgO layers (Fig.~\ref{fig:mgo_geom_sketch}b).
In general, adsorption on top of oxygen is the most energetically favored (Supplementary Fig.~S1a), but we consider both adsorption positions as STM experiments can manipulate the adatoms between
them~\cite{Willke2018a}.
There are three main exceptions to the preferred position.
The energy differences between the bridge and the oxygen-top positions are relatively small ($\Delta E \approx \SI{40}{\milli\electronvolt}$) for Sc and Ti on a single layer of MgO, but the stability of the oxygen-top position greatly increases for the two-layers case ($\Delta E \approx \SI{0.6}{\electronvolt}$).
Ni shows a strong sensitivity to the MgO thickness, switching from the oxygen-top position for the monolayer to the bridge position for the two layers.

The structural trends for the adatoms on the oxygen-top position are as follows.
The adatoms lead to a deformation of the angle $\alpha$ between O and the neighboring Mg (see Fig.~\ref{fig:mgo_geom_sketch}a), which is more pronounced for a single MgO layer (between \SI{150}{\degree} and \SI{165}{\degree}) than for two or three MgO layers (between \SI{160}{\degree} and \SI{175}{\degree}).
This deformation decreases when going from Sc to Cu (see Supplementary Fig.~S2a).
The distance between the 3$d$ adatoms and the O atom below has an average value of \SI{1.90}{\angstrom} for a single layer of MgO, and increases slightly for two and three layers of MgO, with the largest increase for V (see Supplementary Fig.~S2b). This increase for V causes an important effects that will be discussed below.

The geometry of the bridge position is more complex, and in order to characterize it we calculated the O--adatom--O and Mg--adatom--Mg angles, as well as the distances between the adatom and the closest O and Mg atoms (see Supplementary Fig.~S3).
In general, there is a substantial buckling of the square formed by the two Mg and O atoms surrounding the adatom.
The O-adatom-O and Mg-adatom-Mg angles (see Fig.~\ref{fig:mgo_geom_sketch}b) are typically close to each other, and in the range of \SI{60}{\degree} to \SI{80}{\degree} for both one and two layers of MgO.
The buckling becomes more evident by the different distances between the adatom and either Mg or O.
These distances have a large spread for the considered adatoms (\SI{2.3}{\angstrom} to \SI{2.9}{\angstrom}), with Cr having the largest distance to the neighboring atoms and so the smallest buckling.

\begin{figure}[tb]
    \centering
    \includegraphics[width=\textwidth]{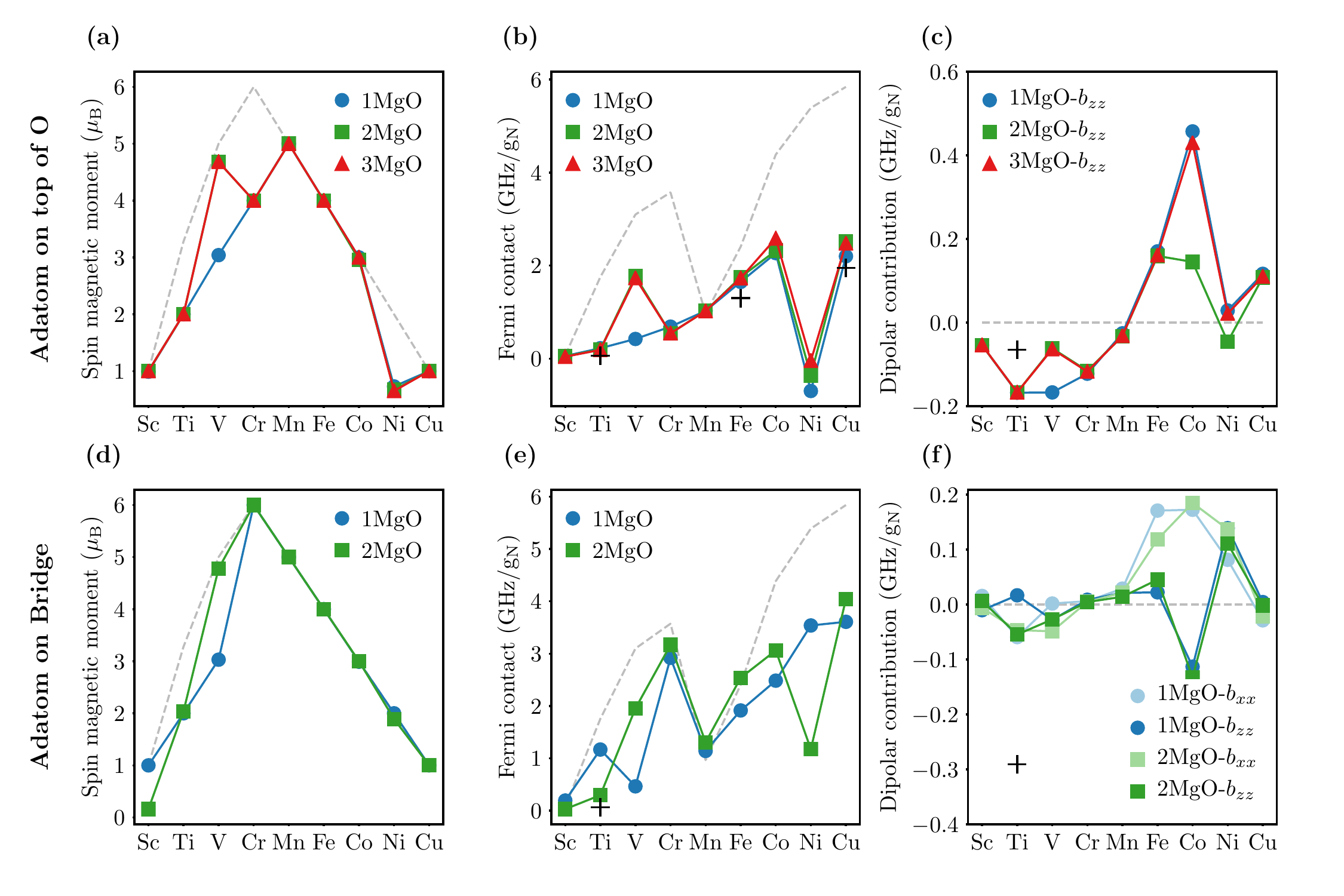}
    \caption{Magnetic properties of adatoms on MgO ultra-thin films, placed on top of oxygen (a--c) and on the bridge position (d--f).
    (a,d) Spin magnetic moment, (b,e) Fermi contact and (c,f) dipolar contributions to the hyperfine interaction.
    The number to the left of MgO in the legend indicates the number of layers in the ultra-thin film. The gray dash line represent the free-atom results. The black crosses in panels (b,c) and (e,f) refers to the experimental results for Fe and Ti from Ref.~\citenum{Willke2018a} and for Cu from Ref.~\citenum{yang2018electrically}.}
    \label{fig:mgo_mag_results}
\end{figure}

\begin{figure}[tb]
    \centering
    \includegraphics[width=\textwidth]{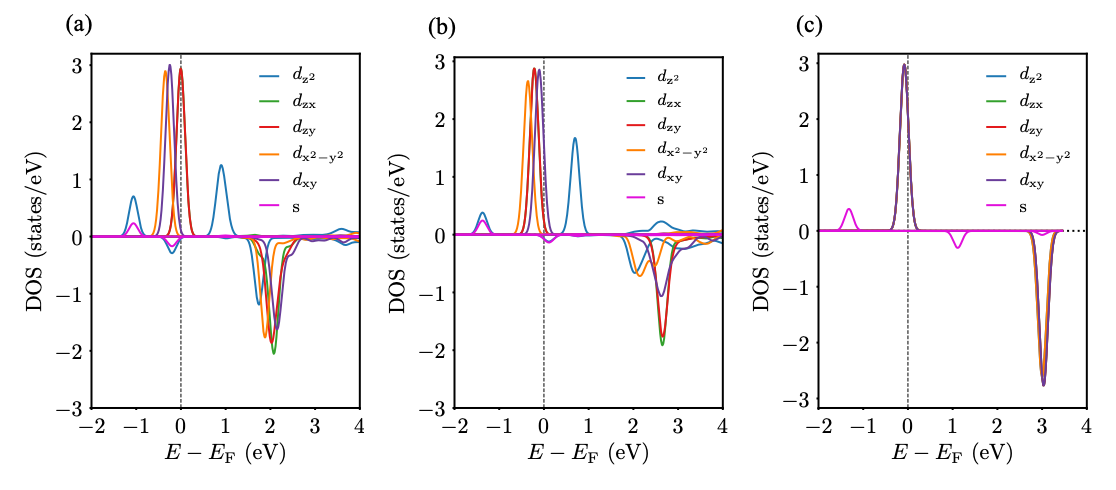}
    \caption{Thickness dependence of the electronic structure of a V adatom on MgO in the oxygen-top position.
    (a) One monolayer of MgO. (b) Two monolayers of MgO. (c) Free atom.}
    \label{fig:mgo_v_results}
\end{figure}

The magnetic properties for the adatoms on the oxygen-top position are illustrated in Fig.~\ref{fig:mgo_mag_results}.
Panel a shows that the spin moments of the adatoms are generally unaffected by the number of MgO layers.
The spin moment of most adatoms follows Hund's first rule and is consistent with a nominal valence of [Ar]$4s^23d^n$.
The Fermi contact contribution to the hyperfine interaction is shown in Fig.~\ref{fig:mgo_mag_results}b.
Our results agree well with the experimental measurements of the Fermi contact for Ti, Fe~\cite{Willke2018a} and Cu~\cite{yang2018electrically} both on top of an oxygen and bridge positions (Ti) of two layers of MgO/Ag(001) surface.
Its overall magnitude is fairly reduced when comparing to the free-atom case (cf.\ Fig.~\ref{fig:atoms_mag}c), consistent with a nominally full $4s$ shell.
The jump in the spin moment of V with increasing MgO thickness occurs due to a change in valence from $4s^23d^3$ (low-spin) to $4s^13d^4$ (high-spin), which also leads to a sudden change in the Fermi contact value. This occurs because both quantities approach the free-atom value as the distance between V and O increase.
This is clearly seen in the modification of the underlying electronic structure shown in Fig.~\ref{fig:mgo_v_results}.
The other notable oddity is Ni, which has a small spin moment of about $0.8\,\mu_\mathrm{B}$ when deposited on top of O and $2.0\,\mu_\mathrm{B}$ when deposited on the bridge position (see Fig.~\ref{fig:atoms_mag}d).
The site-dependent magnetism of Ni on MgO has already been identified theoretically~\cite{PRB1234}.
The Fermi contact contribution lies outside the expected trend of the curve, with a weak and negative value at the oxygen-top position for all MgO thicknesses.
This negative value is due to an antiparallel alignment of the contribution to the spin moment coming from the $s$-electrons with respect to the one of the $d$-electrons.

The presence of the MgO substrate leads to a shape deformation of the spin density that gives rise to a dipolar contribution to the hyperfine interaction, shown in Fig.~\ref{fig:mgo_mag_results}c.
Due to the $C_{4\mathrm{v}}$ symmetry of the oxygen-top position, we show only the dipolar parameter $\mathrm{b}_{zz}$, where $z$ is the direction normal to the MgO layer and coincides with the adatom-oxygen bond orientation.
It follows from Eq.~\eqref{Dipolar} that the spherical part of the spin density does not contribute, which is why the dipolar contribution is not present for the free atoms.
Thus, in contrast to the Fermi contact contribution, we now expect the signal from the $d$ electrons to be more important than the ones from the spherically-symmetric $s$ electrons.

The other important factor is that an ideal half-filled $d$-shell is spherically symmetric and should lead to a vanishing value of the dipolar contribution, as is indeed observed for Mn.
The spin density is formed mostly by $d$-up electrons for Sc--Mn, while Fe--Cu have a dipolar contribution that originates on the $d$-down electrons, as their $d$-up shells are completely filled.
As $\rho_\mathrm{s}(\mathbf{r})=\rho_\uparrow(\mathbf{r})-\rho_\downarrow(\mathbf{r})$, 
it is expected that the sign of the dipolar contribution changes when going from the first to the second group of adatoms --- assuming that the deformation of the spin density is governed by the strong interaction with oxygen and so has the same shape for all adatoms.
The particular example of Co shows that the dipolar contribution is very sensitive to small changes in the atomic environment when going from one to three MgO layers, while the spin moment and the Fermi contact contribution are weakly affected but that only happens to Co, basically.
In the case of V, increasing the thickness of MgO leads to a reduction of the dipolar contribution while both the spin moment and the Fermi contact contribution increase, again indicating that V becomes closer to the free-atom limit.

When the adatoms are placed on the bridge position instead,
 their magnetic properties change substantially.
As seen in Fig.~\ref{fig:mgo_mag_results}d, the spin moments from Cr to Cu match the free-atom values for one and two layers of MgO, while those from Sc to V have a more irregular dependence on the number of MgO layers.
Sc almost loses its magnetic moment for two layers of MgO, Ti has a robust moment of $2\,\mu_\mathrm{B}$ (lower than $3.28\,\mu_\mathrm{B}$ for the free-atom case),
the low- to high-spin transition for V occur also here again.

The situation, however, is more complex than it seems. As revealed by the trends in Fig.~\ref{fig:mgo_mag_results}e, the Fermi contact contribution is overall weaker than in the free-atom case even for those adatoms that have spin moments that match the free-atom values.
Ti and Ni are excellent examples of how the Fermi contact contribution may be sensitive to internal rearrangements of the electronic configuration of the adatom.
Both show strong changes in the value of the Fermi contact when going from one to two MgO layers, even though the total spin moment is unchanged, which indicate changes in the relative contributions to the total spin moment from the $s$ and $d$ electrons.
These are caused by modifications of the buckling of the surrounding Mg and O atoms when the number of layers is increased (see Supplementary Fig.~S3).
Note that now the Fermi contact contribution for Ni is positive.

The dipolar contribution becomes more difficult to interpret for the adatoms on the bridge position, see Fig.~\ref{fig:mgo_mag_results}f.
The main reason is that the local symmetry is now $C_{2\mathrm{v}}$ (instead of $C_{4\mathrm{v}}$ for the oxygen-top position), so the dipolar tensor has two independent parameters instead of one.
As shown in Fig.~\ref{fig:mgo_geom_sketch}b, we set the axes so that $z$ is normal to the MgO plane, with $\mathrm{b}_{zz}$ the corresponding dipolar parameter, and the $x$ axis points to the nearest oxygen, with dipolar parameter $\mathrm{b}_{xx}$.
The trends in the $\mathrm{b}_{zz}$ parameter are qualitatively similar as the previous case but with an overall reduced magnitude, and no substantial dependence on the MgO thickness is found.
A possible explanation for the reduced magnitude is that now the adatom bonds with two oxygen atoms instead of one, and the bond lengths are larger (see Supplementary Fig.~S3). This leads to a weaker deformation of the spin density.
Ti is an exception where the dipolar interaction depends on the number of layers. As found for the Fermi contact part, the rearrangement of the $s$ and $d$ densities should also reflect in a change in its shape, as detected via $\mathrm{b}_{zz}$.
The biaxial character of the deformation of the spin density can be understood by comparing $\mathrm{b}_{zz}$ with $\mathrm{b}_{xx}$ (the other parameter is $\mathrm{b}_{yy} = -\mathrm{b}_{xx} - \mathrm{b}_{zz}$).
Choosing Co as an extreme example, $\mathrm{b}_{xx} \approx -2\mathrm{b}_{zz}$ which implies $\mathrm{b}_{yy} \approx \mathrm{b}_{zz}$, so there is only one primary deformation axis of the spin density, towards the oxygen.
We thus see that a careful study of the dipolar parameters can provide a lot of information about the spatial distribution of the spin density.
There is a large disagreement between our computed and the experimentally measured value of the dipolar contribution for the Ti adatom, which is most likely due to its observed hydrogenated state~\cite{Willke2018a}.

\subsection{Adatoms on 1--2 layers of NaCl and NaF}

To gain further insights into the magnetic properties of the adatoms on ultrathin insulators, we now place them on one and two layers of NaF and NaCl.
These have the same rocksalt structure as MgO, but a larger energy gap.

Given that the structure is the same, one might expect that the relaxed geometries when the adatoms are present would be similar to the case of MgO.
From the energetic point of view, the anion-top position is still the most energetically favored (Supplementary Fig.~S1b,c), with the main exceptions being Sc and Ti for a single layer of NaF and NaCl.
Another important result is that the energetic stability of the anion-top position with respect to the bridge position decreases when going from NaF to NaCl.
We will consider both adsorption positions in the following, as done for MgO.
However, we note that the adatoms could also become substitutional dopants instead~\cite{PhysRevLett.112.026102,chen2014adsorption}.
 
The structural trends for the adatoms on the fluorine-top position of NaF are very similar to those for the oxygen-top position on MgO.
The angle between F and the neighboring Na varies in a slightly larger range than for MgO, while the distances between the adatoms and F is similar to the ones found for MgO (except for Cr, which sits $\sim \SI{0.4}{\angstrom}$ further then the other cases).
The adatoms on the chlorine-top position of NaCl are very sensitive to the number of NaCl layers.
The angle between Cl and the neighboring Na deviates little from the ideal value of \SI{180}{\degree} for two layers of NaCl, while for a single layer the deformation is much stronger (up to \SI{25}{\degree} for Fe-Cu) and less regular.
In general, the distances between the adatoms and Cl are also larger than for MgO or NaF.
More details on the geometry of the systems are given in Supplementary Fig.~S4.

The geometry of the adatoms on the bridge position is also similar to their corresponding ones on MgO (see Supplementary Fig.~S5 and Fig.~S6).
The buckling is somewhat more regular, as the distances between the adatoms and either Na or the respective anions are closer to each other than for MgO, and are on average larger for NaCl than for NaF, as expected.
Cr is once again an outlier, with much larger distances to the neighboring atoms than the other adatoms --- and so the smallest induced buckling of the under-layer. 
The same applies to Mn on two layers of NaCl.
Apart from those exceptions, the angles between each adatom and the neighboring atoms are in the range of \SI{80}{\degree} to \SI{100}{\degree}.

\begin{figure}[tb]
    \centering
    \includegraphics[width=\textwidth]{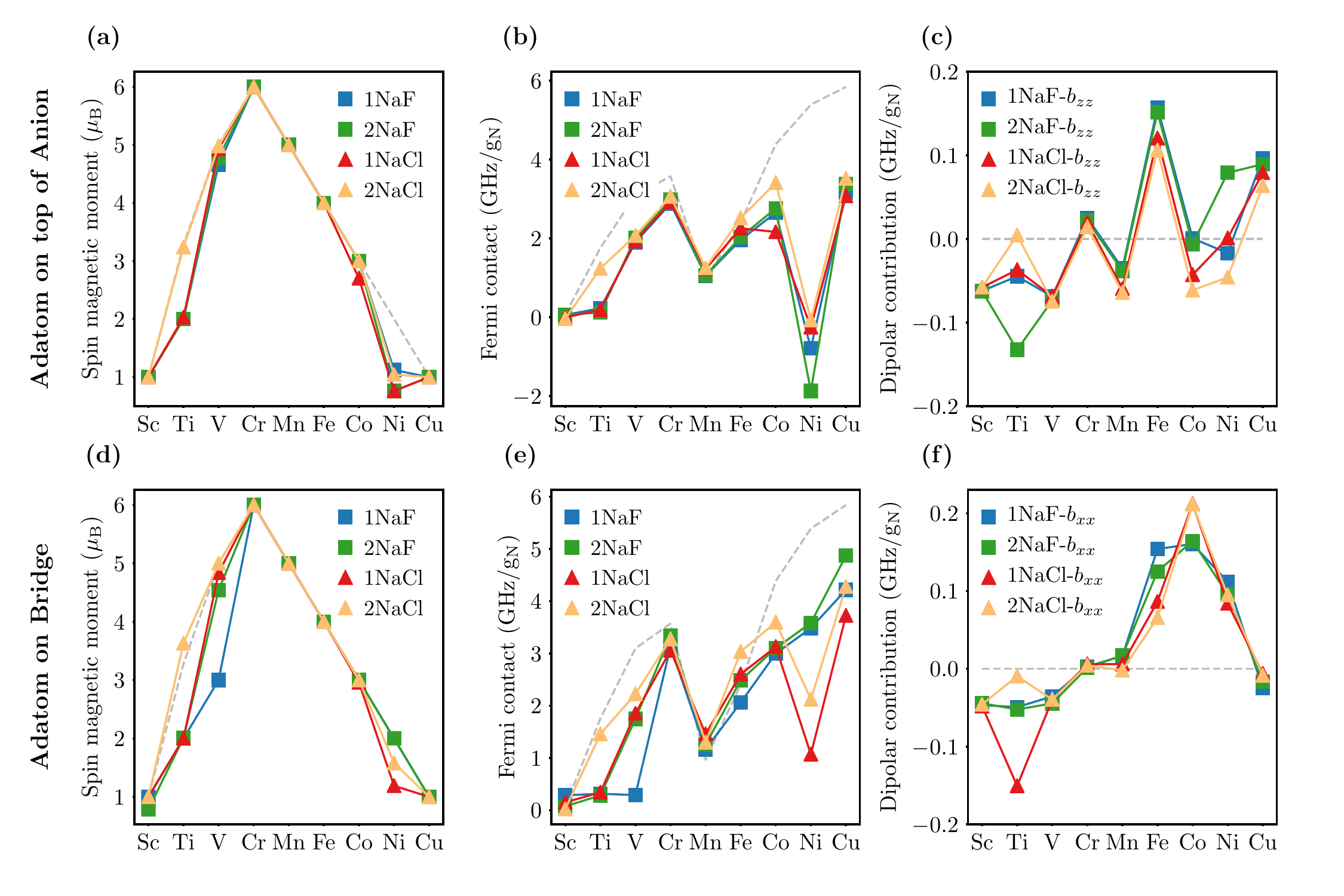}
    \caption{ Magnetic properties of adatoms on NaF and NaCl ultra-thin films, placed on top of anion (a--c) and on the bridge position (d--f).
    (a,d) Spin magnetic moment, (b,e) Fermi contact and (c,f) dipolar contributions to the hyperfine interaction.
    The number to the left of NaF and NaCl in the legend indicates the number of layers in the ultra-thin film. The gray dash line represent the free-atom results.}
    \label{fig:nafcl_mag_results}
\end{figure}

The magnetic properties for the adatoms on the anion-top and on the bridge positions are shown in Fig.~\ref{fig:nafcl_mag_results}, and overall have little dependence on the number of NaF and NaCl layers, with exceptions being Ti, V and Ni.
There is a close parallel between both bridge and anion-top positions to the previous results for adatoms on the bridge position of MgO.
Keeping the noted exceptions in mind, the spin moments tend to assume the free-atom values, and both contributions to the hyperfine interaction have similar trends to the previous cases, as can be identified in Fig.~\ref{fig:all_mag_results}.

Let us turn our attention to particular examples.
Ti on NaF has a low spin, consistent with a valence of $4s^23d^2$, but the dipolar contribution reveals that a reshaping of its spin density takes place when going from one to two layers, but only for the F-top position.
The behavior of Ti on NaCl and of V and Ni on both substrates can be rationalized with the transition between low-spin and high-spin.
Nothing seems to be happening for Fe, Co and Cu if one looks only at their spin moment, but the Fermi contact contribution shows the impact of the changes in the local geometry, mostly for the bridge position.
The dipolar contribution also follows those changes.
While Fe shows similar values of the dipolar parameters for both top and bridge positions, Co and Cu have marked differences.
The dipolar parameters for Co go from large and positive on the top position to small and negative on the bridge position.
In the case of Cu, the behavior of the dipolar parameters combined with the values of the Fermi contact contribution suggest that the spin density becomes more $s$-like on the bridge position, given the increase in magnitude for the Fermi contact and the drop to near zero in the dipolar parameters.\

\subsection{Adatoms on 1 layer of Cu$_2$N and h--BN}

 To further investigate the different behaviours found for the thin insulating layers with rocksalt structure, we turn to insulating materials with different structures, h--BN and Cu$_2$N, commonly used for experiments.
We consider only the nitrogen-top position for both cases, and note that N is less electronegative than O and F and similar to Cl.
Some of these adatoms have been experimentally investigated~\cite{PhysRevLett.109.066101,jacobson2015quantum,gallardo2019large,Hirjibehedin2007,Hirjibehedin2006,otte2008role,Rejali2020,Loth2010,Loth2010a,Loth2012} but without addressing their hyperfine interactions.

The structural trends for the adatoms on the nitrogen-top position of Cu$_2$N  are as follows.
The angle between N and the two diametrically opposite Cu is near the ideal value of \SI{180}{\degree}, with the largest deviations ($\sim \SI{18}{\degree}$) found for V, Cr and Mn, while the distances between the adatoms and N are similar to the ones between the adatoms and the anion on MgO and NaF (see Supplementary Fig.~S7).
The adatoms on the nitrogen-top position of h--BN behave quite differently compared to Cu$_2$N, with the angle between N and B very close to the clean value, and the bonding distances between the adatoms and nitrogen being larger than on Cu$_2$N and similar to those on the top position for NaCl.
These distances are actually so large for the Cr and Mn adatoms that they point to a failure of the PBE exchange-correlation functional in binding these adatoms to the film.
This could be remedied to some extent by obtaining the relaxed geometry with an explicit account of van der Waals interactions (Supplementary Fig.~S7).

\begin{figure}[tb]
    \centering
    \includegraphics[width=\textwidth]{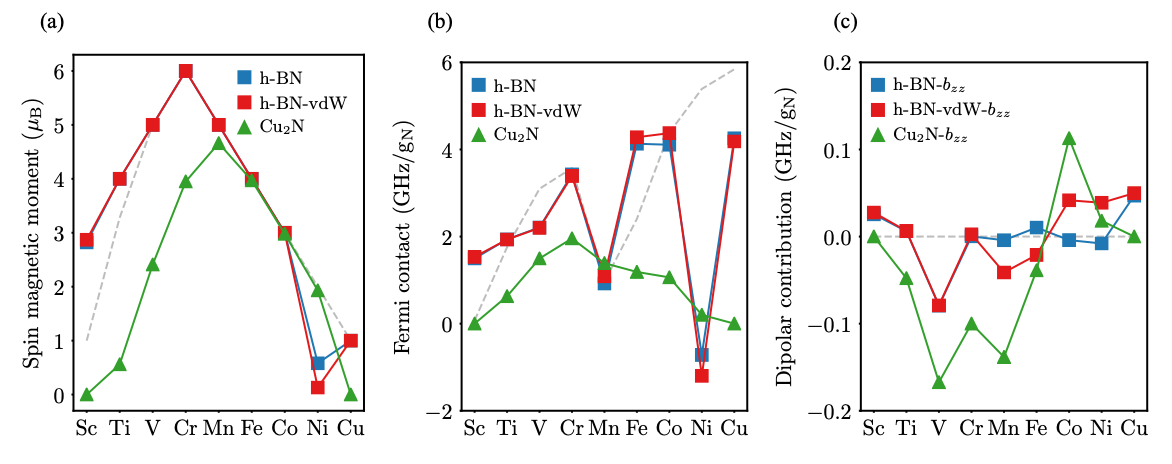}
    \caption{Magnetic properties of top-stacked adatoms on Cu$_2$N and h--BN.
    (a) Spin magnetic moment, (b) Fermi contact and (c) dipolar contributions to the hyperfine interaction. The gray dash line represent the free-atom results.}
    \label{fig:cu2bn_mag_results}
\end{figure}

The magnetic properties for the adatoms on the nitrogen-top position of h--BN and Cu$_2$N are shown in Fig.~\ref{fig:cu2bn_mag_results}.
Starting with h--BN, the magnetic properties of most of the adatoms are weakly influenced by the vdW correction.
At first glance, it seems that the adatoms adopt the free-atom values for the spin moments (except Ni), but a closer inspection reveals that Sc and Ti have an even higher spin than as free atoms.
This is due to their unusual electronic configuration, [Ar]$4s^13d^{n+1}$, that arises from the $s$-down energy level rising above the $d$-up energy levels, a behavior already found in a theoretical previous study~\cite{PhysRevB.82.045407}.
This explains the overall large values for the Fermi contact and the weak ones for the dipolar contributions to the hyperfine interaction.
The negative values of the Fermi contact parameters for Ni were also found for the top positions of the previously discussed layers (clearly seen in Fig.~\ref{fig:all_mag_results}a).
The increased values of the dipolar parameters for Mn, Fe and Co when the van der Waals correction is included can be explained by the shortening of the adatom-nitrogen distances.
Our findings are in broad agreement with those of Ref.~\cite{PhysRevB.82.045407}, which also considered the impact of corrugation and a metal substrate.

Lastly, we discuss the magnetic properties for the Cu$_2$N case.
The spin moment vanishes for Sc and Cu, weakens for Ti, and strengthens for Ni, being overall far from the free atom limit.
Computationally, Cu$_2$N is found to have a very small band gap, so that the magnetically active orbitals of the adatoms are hybridizing with the conduction and/or valence states of the layer, instead of lying in a large energy gap as in the cases for all the previously studied systems.
There is no longer a strong contribution to the spin moment from the $s$-orbitals, which leads to the weakest values found for the Fermi contact parameters.
The large value of the dipolar parameters for V, Cr, Mn and Co can also be assigned to the stronger interaction with the electronic states of the layer.

\section{Discussion}

We presented a comprehensive set of ab initio calculations for the hyperfine interactions of magnetic 3$d$ adatoms on different ultra-thin insulating layers (MgO, NaF, NaCl, h--BN and Cu$_2$N) of varying thicknesses.
We also obtained structural and electronic properties and the spin magnetic moments, which we used to interpret the hyperfine interactions.
The free-atom calculations provide a clear picture for the origin of the hyperfine interaction and serves as a useful reference against which we could compare and contrast the behavior of the same atoms on different films.
We quantitatively established that the hyperfine interactions are very sensitive to the strength of the interaction between the adatom and the thin film and to the local arrangement of the atoms, and can provide information about the electronic structure and magnetism of these adatoms that cannot be experimentally determined by other means.

The basic principles for understanding the hyperfine interactions were illustrated with the reference atomic calculations.
The dominant contribution is called Fermi contact and is determined by the spin density at the nuclear position.
As this arises solely from the $s$ electrons, it allows us to separate the total spin moment into $s$ and $d$ contributions, and to follow the relative energies of the $s$ and $d$ levels.
The second contribution is called dipolar, as it represents the dipole-dipole interaction between the electronic and nuclear spins.
This is only present if the electronic spin density is nonspherical, and so is absent for the free atoms.
A summary of our findings is given in Fig.~\ref{fig:all_mag_results} for both contributions in all calculated substrates.

The hyperfine interactions are experimentally measurable quantities, but only a few measurements are available for adatoms on MgO.
In order to obtain systematic physical insights and provide guidance and reference values for future experiments, we decided to map the trends in these interactions and related properties for adatoms on the anion-top and bridge positions of rocksalt ultra-thin insulating layers: MgO, NaF and NaCl.
Adsorption on the anion-top position was generally found to be the most energetically favored, but we considered both positions as it is experimentally possible to manipulate the adatoms between them. 
The structural trends for the adatoms on NaF are very similar to MgO, while on NaCl there is a stronger dependence on the number of layers.
On the top position, the distance between the adatoms and Cl is substantially larger than for the O and F cases.
The buckling of the atoms surrounding the adatom on the bridge position of NaF and NaCl is somewhat more regular than on MgO.

The behavior of the spin magnetic moments can be understood as switching between the Hund's first rule for the $d$ electrons (maximizing the number of $d$-up minus number of $d$-down electrons) and fully polarizing the $s$ electrons (which is seen for V and Cr in most cases).
This is explained by changes in the energetic alignment between the $s$ and $d$ levels, and can lead to transitions between two total spin values when changing the number of layers of the film.
The magnitude of the Fermi contact parameter is strongly reduced in comparison with the free-atom case (see Fig.~\ref{fig:all_mag_results}a), which can signal either the reduction in spin polarization of the $s$ electrons or the reduction of their spatial localization near the nucleus of the adatom, due to bond formation with the atoms in the film.
Overall, the larger the Fermi contact parameter is, the closer the adatom is to the free-atom limit.
The Fermi contact parameter can also trace internal rearrangements of the electronic structure that do not change the total spin, as found for Ti and Ni on the bridge position of MgO.
The dipolar contribution to the hyperfine interaction is mostly sensitive to the spin density arising from the $d$ electrons, and how it is deformed by interaction with the substrate.
The dipolar parameters $\mathrm{b}_{zz}$ (top position) and  $\mathrm{b}_{xx}$ (bridge position) are related to the interaction between the adatom and the anions.
Their sign correlates well with whether the number of $d$ electrons is lower (negative) or higher (positive) than five, while the magnitude quantifies the non-sphericity of the spin density, and is related to changes in the local bonding geometry.
Comparing the hyperfine interactions on different films, there is a close parallel between both bridge and anion-top positions on NaCl and NaF and what was found for adatoms on the bridge position of MgO, while the top position on MgO has a more distinct behavior.
The overall magnitude of the Fermi contact contribution for the top position is somewhat stronger on NaF and NaCl than on MgO (see Fig.~\ref{fig:all_mag_results}a), which is caused by the larger distance between the adatoms and the anion.

To gain further insight into the interaction between the adatoms and the anions on the top position, we expanded our dataset to include single layers of Cu$_2$N and h--BN.
These are more covalently-bonded materials, while the rocksalt compounds are more ionic, which leads to several differences.
The adatom-anion distances on Cu$_2$N are similar to those on MgO and NaF, while h--BN compares to NaCl.
These large distances in case of h--BN result in the adatoms adopting the free-atom values for the spin moments (except Ni) or even higher, as seen for Sc and Ti, due to their unusual electronic configuration that arises from the $s$-down energy level rising above the $d$-up energy levels.
This explains the overall large values for the Fermi contact and the weak ones for the dipolar contributions to the hyperfine interaction compared to the other studied films (see Fig.~\ref{fig:all_mag_results}).
Despite the similar adatom-anion distances, the magnetic properties of the adatoms on Cu$_2$N are quite different from those on NaF and MgO.
This is due to the electronic structure of Cu$_2$N, which is found computationally to have a small band gap, so that the magnetically active orbitals of the adatoms are hybridizing with the conduction and/or valence states of the layer, instead of lying in a large energy gap as was the case for all the previously studied systems.
We also found the weakest values for the Fermi contact parameters, as there is no longer a strong contribution to the spin moment from the $s$-orbitals.

We conclude that the most important properties from which all else follows for our magnetic adatoms on thin insulating layers are (i) the energetic alignment between the $s$ and the $d$ levels, and (ii) how strong is the interaction between the magnetic orbitals and those of the substrate.
These are quite challenging to compute accurately with standard exchange-correlation functionals, but we believe that our uncovered trends and associated discussion should hold in general even if the specific values could be refined.
As it is important to treat both the $s$ and $d$ orbitals of the adatoms on equal footing (given that the hyperfine interactions depend on both) and also to accurately describe the valence and conduction states of the thin film (for instance to get an improved band gap), future studies should consider adopting hybrid functionals or other nonlocal exchange-correlation corrections.
Another relevant aspect that we have not considered in the present study is the role of the spin-orbit interaction.
This should be studied in conjunction with the aforementioned improved description of the electronic structure, so as to restore Hund's rule for the orbital magnetic moment, which is then coupled to the spin moment via the spin-orbit interaction.
Lastly, on the structural level, van der Waals interactions can lead to strong modifications in a few cases, as we already verified for the case of h--BN.

\section{Methods}
Structural, electronic and magnetic calculations were carried out within the framework of density functional theory (DFT).
We used Quantum Espresso~\cite{Giannozzi2009,Giannozzi2017} with pseudopotentials from the PSLibrary~\cite{DalCorso2014} and the projector augmented wave (PAW) method ~\cite{PhysRevB.50.17953}.
Exchange and correlation effects were treated in the generalized gradient approximation of Perdew, Burke and Ernzerhof (PBE)~\cite{PhysRevLett.77.3865}. 
The hyperfine parameters in Eqs.~\ref{Fermi} and~\ref{Dipolar} were computed with the GIPAW module of Quantum Espresso, which implements the theory developed by Pickard and Mauri~\cite{Pickard2001}.
For all calculations, the kinetic energy cutoff for the wavefunctions and for the charge density were set to \SI{90}{\rydberg} and \SI{720}{\rydberg}, respectively.
The Brillouin zone integrations were performed with a Gaussian smearing of width \SI{0.01}{\rydberg}.

For the calculations of isolated atoms, we employed cubic periodic cells with length \SI{10}{\angstrom}, in order to minimize interactions between periodic replicas of the atom, and $\Gamma$-point sampling of the Brillouin zone.

In order to determine the theoretical lattice constants, we constructed monolayer, bilayer and trilayer geometries with 4 atoms per layer for MgO, NaF and NaCl,
and 3 and 2 atoms for monolayers of $\mathrm{Cu}_{2}$N and h--BN, respectively.
In all cases, we included a vacuum thickness equivalent to 9 layers of the respective materials, and the k-mesh was set to $12\times12\times1$.
The results are collected in Table~\ref{tab:alat}.

\begin{table}[!h]
\begin{ruledtabular}
\begin{tabular}{c c c c c}
 Material & Experimental lattice constants & Monolayers & Bilayers & Trilayers \\
 \hline
 MgO  &  4.2113 \cite{boiocchi2001crystal} (bulk) & 4.0530 & 4.1466  & 4.1846  \\
 NaCl & 5.6418  \cite{fontana2011characterization} (bulk) & 5.6710 & 5.5941 & --- \\
 NaF &  4.6324 \cite{rao1990structural} (bulk) & 4.6555  & 4.6109 & --- \\
 Cu$_2$N &  3.814 \cite{zachwieja1990ammonothermalsynthese} (monolayer) & 3.6637 & --- & ---    \\
 h--BN & 2.50 \cite{doi:10.1021/nl9011497}(monolayer) & 2.5153 &  --- & --- \\
\end{tabular}
\end{ruledtabular}
\caption{\label{tab:alat} Experimental vs.\ theoretical lattice constants for the considered systems.}
\end{table}

To accommodate the adatoms on MgO, NaF and NaCl, we then set up $3\times3$ supercells with one of the 3d adatoms either on top of the anion or in the bridge position, as shown in Fig.~\ref{fig:mgo_geom_sketch}), with a consistent reduction of the k-mesh $4\times4\times1$.
The supercells contained 37, 73 and 109 atoms for monolayers, bilayers and trilayers, respectively, and a vacuum thickness equivalent to 9 layers of the respective materials, as before.
The cell dimensions were kept fixed while all atomic positions were allowed to fully relax.
A similar procedure was followed for the 3d adatoms on top of nitrogen for the monolayers of $\mathrm{Cu}_{2}$N  (28 atoms) and h--BN (19 atoms), with a vacuum thickness of \SI{16.5}{\angstrom} and \SI{16.3}{\angstrom}, respectively.
For the relaxation distance of 3d adatoms on monolayer h--BN, we additionally take into account weak interactions --- the van der Waals correction (DFT-D) --- to improve the description of the binding energy.

\section{Data availability}
The data that support the findings of this study are available from the corresponding author upon reasonable request.

\section{Acknowledgements}
This work was supported by the Federal Ministry of Education and Research of Germany in the framework
of the Palestinian-German Science Bridge (BMBF grant number 01DH16027),and the European Research Council (ERC) under the European Union's Horizon 2020 research and innovation program (ERC-consolidator grant 681405 -- DYNASORE). 
We acknowledge the computing time granted by the JARA-HPC Vergabegremium and VSR commission on the supercomputer JURECA at Forschungszentrum Jülich~\cite{jureca} and RWTH Aachen University under project jara0189.

\section{Author contributions}
S.L. initiated, designed and supervised the project. S.S. performed the first-principles calculations under the supervision of M.d.S.D.   
All the authors discussed the obtained results and contributed to writing and revising the manuscript.

\section{Competing interests}
The authors declare no competing interests.

\end{document}